\documentclass[prd, twocolumn, superscriptaddress, nofootinbib]{revtex4}
\usepackage{graphicx, epsfig}

\begin{document}
\newcommand {\ds}{\displaystyle}
\newcommand {\tg}{\tilde{\gamma}}
\def\hinvmpc{h^{-1}\,{\rm Mpc}}
\def\hmpcinv{h\,{\rm Mpc}^{-1}}

\title{Nulling Tomography with Weak Gravitational Lensing}

\author{Dragan Huterer}
\affiliation{Kavli Institute for Cosmological Physics
and Astronomy and Astrophysics Department,
University of Chicago, Chicago, IL~~60637}

\author{Martin White}
\affiliation{Departments of Astronomy and Physics,
University of California, Berkeley, CA 94720} 

\begin{abstract}
We explore several strategies of eliminating (or nulling) the small-scale
information in weak lensing convergence power spectrum measurements in order to
protect against undesirable effects, for example the effects of baryonic
cooling and pressure forces on the distribution of large-scale structures. We
selectively throw out the small-scale information in the convergence power
spectrum that is most sensitive to the unwanted bias, while trying to retain
most of the sensitivity to cosmological parameters.  The strategies are
effective in the difficult but realistic situations when we are able to guess
the form of the contaminating effect only approximately.  However, we also find
that the simplest scheme of simply not using information from the largest
multipoles works about as well as the proposed techniques in most, although not
all, realistic cases.  We advocate further exploration of nulling techniques
and believe that they will find important applications in the weak lensing data
mining.
\end{abstract}

\maketitle
\section{Introduction}

Progress in cosmology in the last decade has been dramatic, and there is reason
to believe that we will make further advances on key questions in the coming
years.  Key to this progress is the ability to confront precise data with
highly accurate theoretical predictions.  Gravitational lensing illustrates
both the potentials and difficulties of this era of cosmological research.  In
principle, lensing can constrain the properties of the dark energy causing the
accelerated expansion of the universe and constrain both the cosmic geometry
and the growth of large scale structure (e.g.\ \cite{tomography, Hui,
Huterer_2002, Hu_tomo_2, wl_space_III, Abazajian, Song_Knox, Takada_Jain,
Ishak}).  But one of the main challenges for future weak lensing surveys will be
controlling the theoretical systematic errors involved in predicting the
lensing signal at small, nonlinear scales of $\ll 1$ degree.  These errors
include numerical artifacts in computing the non-linear power spectrum (see
e.g.\ \cite{Vale_White, White_Vale, LosAlamos, Huterer_Takada}) 
and complex baryonic processes that are difficult to model
accurately \cite{White_baryon, Zhan_Knox} and make semi-analytic predictions
of nonlinear power unreliable.

Many of the difficulties in making theoretical predictions come in at small
physical scales.  In this paper we explore one approach to protect against
theoretical biases at small scales.  We suggest selectively throwing out the
small-scale information and ``nulling'' the bias while at the same time
retaining useful cosmological information.  What makes us optimistic in this
regard is that the information contained in the weak lensing convergence
estimated in different redshift slices is strongly overlapping, essentially
because of the width of the lensing kernel.  Therefore, dropping a reasonably
small subset of the tomographic information leads to small degradations in
overall cosmological constraints.  This has been used by Takada \& White
\cite{Takada_White} to propose dropping the autospectra in order to protect
against unwanted biases due to intrinsic alignments of galaxies.  We take this
general idea further and propose identifying specific linear combination of the
cross-power spectra that are most sensitive to a given nonlinear effect, and
then dropping them in the analysis in order to null out the effect while
ideally preserving most of the sensitivity to cosmological parameters.

In the next Section we describe the fiducial assumptions about the survey and
small-scale bias. In Sec.~\ref{sec:1pt} we describe and explore several nulling
tomography techniques. We discuss the results and comment on future prospects
in Sec.~\ref{sec:discussion}.

\section{Methodology and small-scale bias model}\label{sec:method}

Assume a weak lensing survey with the ability to divide source galaxies
in $N_s$ redshift bins.  Let $P_{ij}^{\kappa}(\ell)$ be the convergence
cross-power spectrum in $i$th and $j$th tomographic bin at a fixed
multipole $\ell$ --- for definitions and details, see e.g.\ Ref.~\cite{Huterer_2002}. 
The observed convergence power is
\begin{equation}
C^{\kappa}_{ij}(\ell)=P_{ij}^{\kappa}(\ell) + 
\delta_{ij} {\langle \gamma_{\rm int}^2\rangle \over n_i}
\label{eq:C_obs}
\end{equation}
\noindent where $\langle\gamma_{\rm int}^2\rangle^{1/2}$ is the rms intrinsic
shear in each component which we assume to be equal to $0.22$, and $\bar{n}_i$
is the average number of galaxies in the $i$th redshift bin per steradian.

For definiteness we assume a SNAP-type survey \cite{SNAP} covering 1000 sq.\
deg.\
with the galaxy distribution of the form $n(z)\propto z^2\exp(-z/z_0)$ that
peaks at $2z_0=1.0$.  We assume 100 usable galaxies per arcmin$^2$. We use
information from the wide range of scales corresponding to multipoles $100\leq
\ell\leq 10,000$.  We have a set of six cosmological parameters: energy density
and equation of state of dark energy $\Omega_{\rm DE}$ and $w$, spectral index
$n$, matter and baryon physical densities $\Omega_M h^2$ and $\Omega_B h^2$,
and the amplitude of mass fluctuations $\sigma_8$. Throughout we assume a flat
universe.  For our fiducial results, we let these six parameters vary without
any priors since the fiducial survey alone is powerful enough to determine the
cosmological parameters to a good accuracy.  The cosmological constraints can
be computed using the standard Fisher matrix formalism.


To be definitive, let us now consider systematic bias in the matter
power spectrum $P(k)$ of the form
\begin{equation}
P(k, z)=\bar{P}(k, z)\left [ 1 + a\,f(k) \right ]
\label{eq:Pk_bias}
\end{equation}
\noindent where $\bar{P}(k, z)$ is the true, unbiased power spectrum, and $a$
is the coefficient that we use to ``turn on'' the bias $f(k)$ whose functional
form is, we assume for a moment, precisely known. [Note that realistic biases
$f(k)$ may be redshift dependent as well, perhaps increasing at late times
during the nonlinear structure formation, but we ignore this issue since it is
unimportant in the following discussion.]  In order to use weak lensing to
measure the cosmological parameters, we are forced to fit for the parameter $a$
as well, otherwise our results will be biased.  {\it Our goal here is to
minimize the sensitivity of the data on parameter $a$}, since the results would
thus become less sensitive to the particular bias in the power spectrum from
Eq.~(\ref{eq:Pk_bias}). At the same time, we hope not to significantly weaken
the sensitivity to other cosmological parameters, such as the equation of state
of dark energy $w$.

\begin{figure*}[!t]
\vspace{-1cm}
\epsfig{file=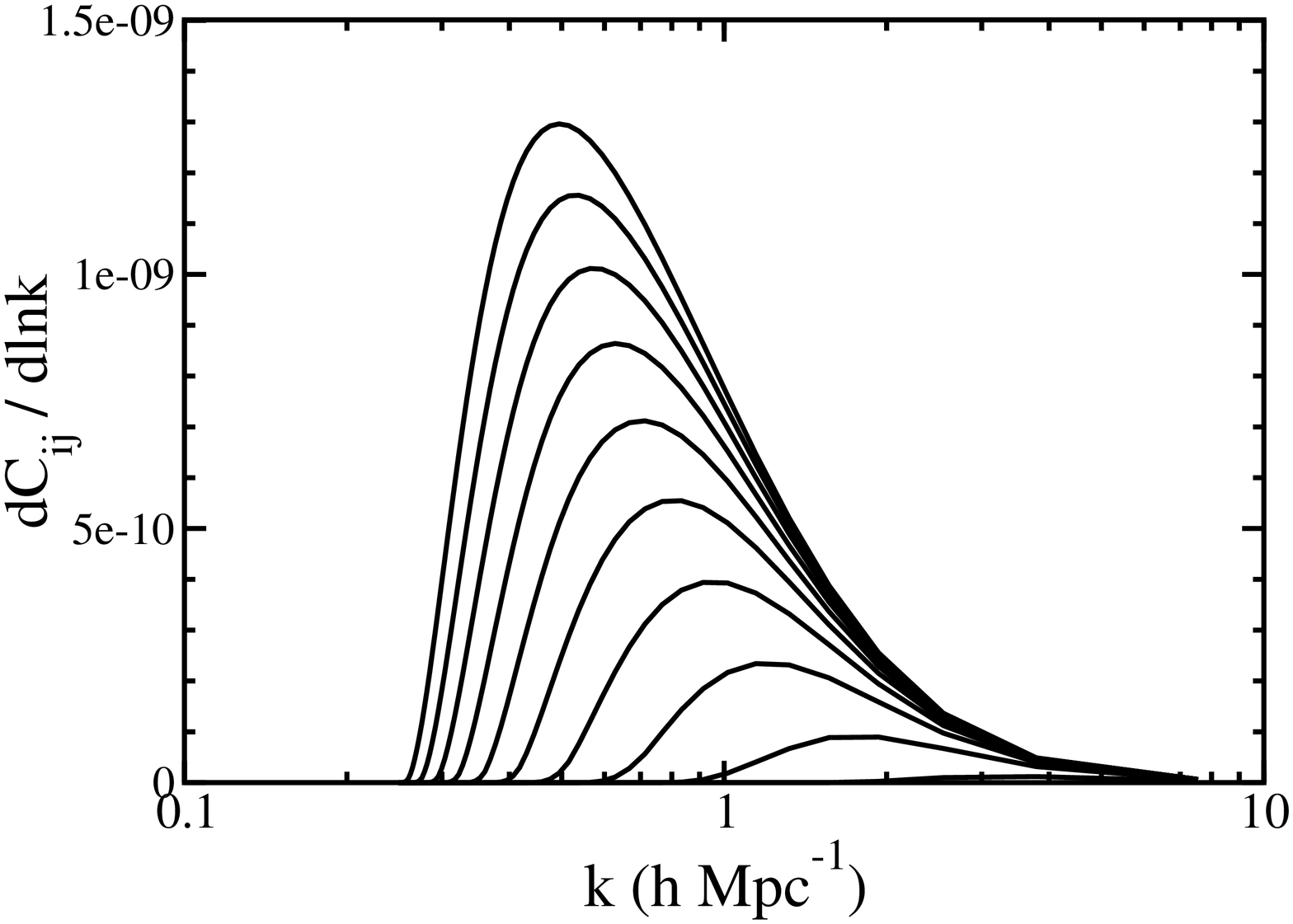,width=3.0in,height=3.4in,angle=-90}\hspace{-0.2cm}
\epsfig{file=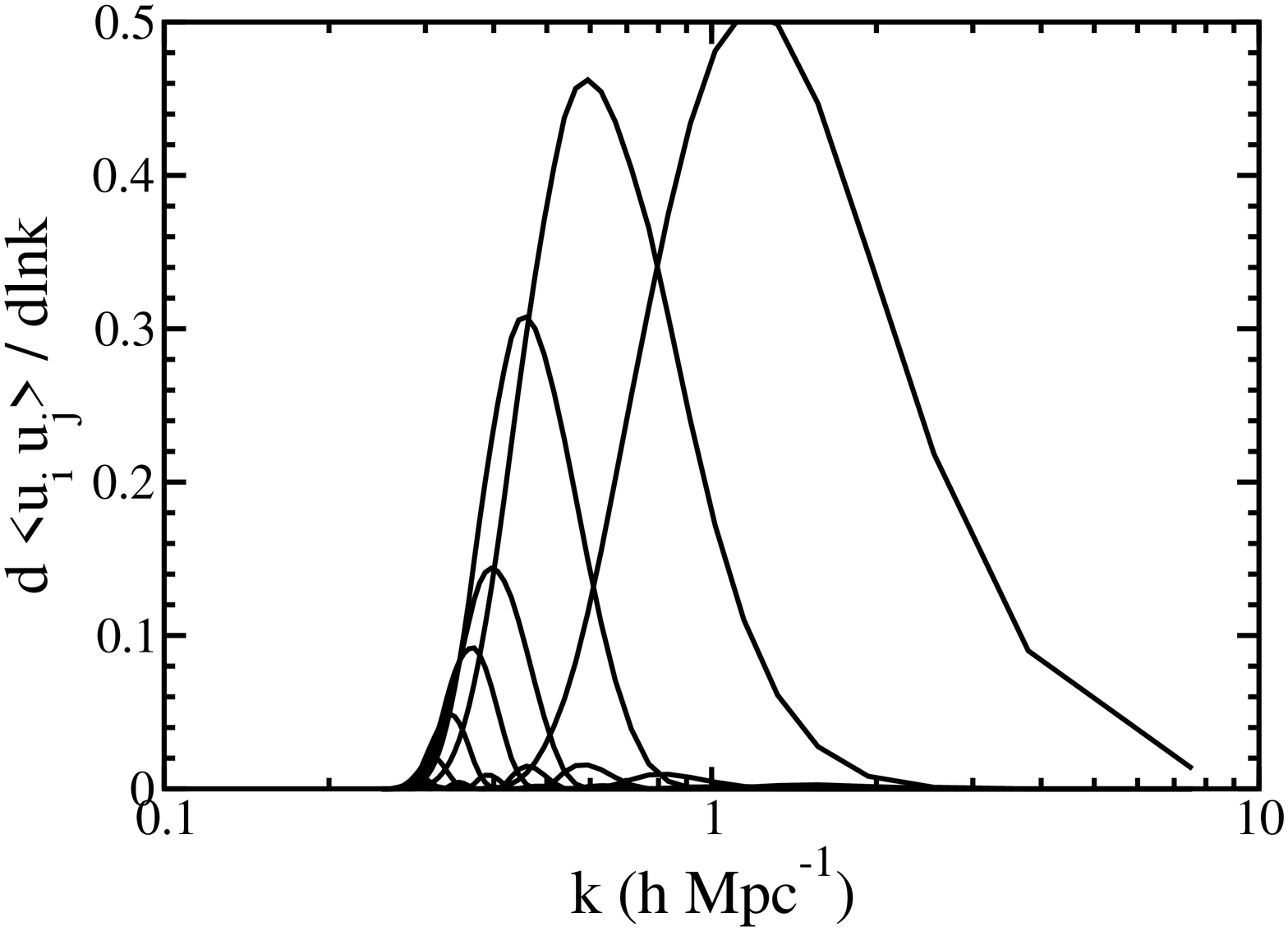, width=3.0in, height=3.4in, angle=-90}
\vspace{-0.3cm}
\caption{Derivatives $dC_{ij}^{\kappa}/d\ln k$ (left panel) and $d\langle u_i
u_j\rangle /d\ln k$ (right panel) at multipole $\ell=1100$. Here
$C_{ij}^{\kappa}$ are the original cross-power spectra and $\langle u_i
u_j\rangle$ are the covariance of the new combinations of shear, defined in
Eq.~(\ref{eq:Cnew}).  Note the coverage in $k$ for the new eigenmodes; the ones
that have most of their weight at high wavenumbers are most sensitive to the
parameter $a$. The overall normalization in each panel is arbitrary, but the
relative size of the modes is not. }
\label{fig:weights_1pt}
\end{figure*}

For the bias in the matter power spectrum, our fiducial model is as in
Eq.~(\ref{eq:Pk_bias}) with $a=1.0$ and $f(k)= (k/k_*)^{\alpha}$. This model
phenomenologically describes the baryonic effects if we further set $\alpha\sim
3$ and $k_*\sim 10\hmpcinv$ \cite{White_baryon, LosAlamos} and these are
the values we use, although we explored a range of other values and found similar
results. Finally, to compute the bias in the cosmological parameters due to the
small-scale bias, we follow the standard formalism that uses the Fisher matrix
described in e.g.\ Ref.~\cite{Huterer_Takada}.

\section{1-point nulling}\label{sec:1pt}

Let us first consider using $N_s$ specific linear combinations of shears in
individual redshift bins, specifically $\sum_{j=1}^{N_s}
Q_{ij}\gamma_j(\mbox{\boldmath$\theta$})$, where the coefficients $Q_{ij}$ are
independent of the angular position of the galaxy. In other words, each linear
combination gives equal weight to all galaxies in the $j$th redshift bin. Our
goal is to find specific linear combinations that are most sensitive to the
nuisance parameter $a$ and simply throw them out. The nice feature of this problem
is that it is tractable analytically.

Let us fix the multipole $\ell$ and define $\tg_{(\ell, m), i}\equiv a_{(\ell, m)
i}$, where $a_{(\ell, m), i}$ is the spherical harmonic
coefficient of the shear map in redshift bin $i$. The only things
we really need to know about $\tilde{\gamma}_{(\ell, m), i}$ are its mean and variance
\begin{equation}
\langle \tg_{(\ell, m), i} \rangle     =0, \,\,\,\,\,\,
\langle \tg_{(\ell, m), i} \tg_{(\ell', m'), j}\rangle = 
\delta_{\ell \ell'}\delta_{mm'} C^{\kappa}_{ij}(\ell),
\label{eq:gamma_var}
\end{equation}
\noindent Now consider the linear combinations
\begin{equation}
u_i = \sum_{j=1}^{N_s} Q_{ij}\tg_{(\ell, m), j}.
\label{eq:u}
\end{equation}
\noindent where hereafter we drop the subscript $(\ell, m)$ from $u_i$.
The covariance matrix of those combinations is
\begin{equation}
\langle u_i u_j\rangle = \sum_{k, l} Q_{ik}Q_{jl}C_{kl} = {\bf q}^{(i)T}\, {\bf C}\,
{\bf q}^{(j)}
\label{eq:Cnew}
\vspace{-0.3cm}
\end{equation}
\noindent where the column vector ${\bf q}^{(i)}$ is defined as the i$^{th}$
{\it row} of the matrix ${\bf Q}$, and the matrix ${\bf C}$ has elements
$C^{\kappa}_{kl}=\langle \tilde{\gamma}_{(\ell, m), k} \tilde{\gamma}_{(\ell,
m), l}\rangle$. Say we only had a measurement of a {\it single} $u_i\equiv
u$. The likelihood for $u$ (in the Fisher matrix approximation) is gaussian
\begin{equation}
\mathcal{L}\propto {1\over \sqrt{R}} \exp{\left ({-u^2\over 2R}\right)}.
\end{equation} 
\noindent where $\langle u^2\rangle \equiv R = {\bf q}^T\,{\bf C}  \,{\bf q}.$
and we suppressed the index $i$ on ${\bf q}$ as well.
Then we want to determine the ${\bf q}$ by maximizing the error in
$a$, or minimizing (after a bit of algebra)
\begin{equation}
F_{aa}
\equiv \left < -{\partial^2 \over \partial a^2}\ln \mathcal{L}\right >
= {1\over 2R^2} \left ({\partial R\over \partial a}\right )^2 
\label{eq:Faa}.
\end{equation}
We now need to solve the following problem: find the column vector ${\bf q}$
that maximizes $F_{aa}$ in Eq.~(\ref{eq:Faa}),
given the $N_s\times N_s$ covariance matrix ${\bf
C}$. Fortunately, this problem is well known and can be solved using standard
techniques (e.g.\ \cite{TTH,Watkins}).  The solution is to consider 
\begin{equation}
\left ( {\bf L}^{-1} {\partial {\bf C}\over \partial a} ({\bf L}^{-1})^T
\right ) \left ( \bf {L}^T {\bf q}\right ) = 
\lambda  \left ( \bf {L}^T {\bf q}\right ). 
\label{eqn:evalue}
\end{equation}
\noindent where ${\bf L}$ is a lower triangular matrix satisfying
${\bf C}=\bf{L L^T}$.
Eq.~(\ref{eqn:evalue}) is an eigenvalue-eigenvector relation for
${\bf L}^T {\bf q}$ which we can solve to obtain ${\bf q}$.  It can
easily be shown that the error in $a$ from measurement of any single
combination $u_i$ is $\sqrt{2}/|\lambda_i|$, and that $\langle u_i u_j\rangle =
\delta_{ij}$. Therefore, we get the whole spectrum of uncorrelated combinations
of shear ordered by their sensitivity to the small-scale parameter $a$. Again,
recall that we need to repeat the same procedure at every multipole $\ell$ (or,
multipole bin centered at $\ell$).

We find that the Fisher information on the parameter $a$ is heavily dominated
by one (or a few) particular combinations of shears. For example, for the
multipole bin centered at $\ell=1100$ and 4-bin tomography, the eigenvalues
$\lambda$ are: $7.0\cdot 10^{-6}$, $4.2\cdot 10^{-5}$, $3.3\cdot 10^{-4}$, and
$1.3\cdot 10^{-2}$.  Clearly, by far the most information about the nuisance
parameter $a$ is carried in the eigenvector corresponding to the fourth
eigenvalue. This eigenvector is (after adjusting the unimportant overall
normalization for clarity)
$\tilde{\gamma}_1-5.3\tilde{\gamma}_2+10.2\tilde{\gamma}_3 +
5.8\tilde{\gamma}_4$. Therefore, this combination of shears at $\ell=1100$ is
most sensitive to the parameter $a$ and would be the one to drop.

Figure \ref{fig:weights_1pt} shows $dC_{ij}^{\kappa}/d\ln k$ and $d\langle u_iu_j\rangle /d\ln k$ at
multipole $\ell=1100$, now for a 10-bin tomography, where $C_{ij}^{\kappa}$ are the
original cross-power spectra and $\langle u_iu_j\rangle$ are the covariances of the 
linear combinations of shears $u_i$ defined in Eq.~(\ref{eq:u}).
It is clear that the covariances of the $u_i$ have a more differentiated
wavenumber dependence than the original $C_{ij}^{\kappa}$.  However, because of
the width of the gravitational lensing kernel, the weights $d\langle
u_iu_j\rangle /d\ln k$, shown in right panel, are still strongly overlapping.

First, consider the fiducial case where we keep all information in the survey,
but add a single new parameter $a$ describing the small-scale effect. If the
form of the small-scale effect is known exactly, we find that the additional
degeneracy introduced by the new parameter $a$ is negligible, and marginalized
errors in cosmological parameters increase by only a few percent relative to
the case when $a$ is fixed. This is not surprising, as the parameter $a$ enters
the observables in a different way than the cosmological parameters, and the
fiducial survey is powerful enough to determine them all without substantial
loss of accuracy.

Let us now drop the combination $u_i$ that is the most sensitive to parameter
$a$.  We find that the marginalized error on the parameter $a$ increases by
several orders of magnitude while increasing the cosmological parameter
accuracies only by several tens of percent.  While this sounds like fantastic
news, we should remember that we optimistically assumed that we were able to
{\it exactly} guess --- and parametrize --- the form of the small-scale effect
$f(k)$. In reality, we will be able to guess $f(k)$ only approximately, and the
difference between the true and guessed $f(k)$ will lead to biases in the
measured cosmological parameters.  The real figure of merit is the ratio of the
bias due to the incorrectly guessed $f(k)$ and the fiducial accuracy in any
given cosmological parameter.  Clearly, our goal should be to strike balance
between the systematic bias and the statistical precision, minimizing the
former while not increasing the latter by more than a few tens of percent.

Table \ref{tab:degrade_1pt} shows the errors and biases in $w$ when between
zero and two combinations of shear, out of ten total, were dropped, and
alternatively, when multipoles $\ell>\ell_{\rm max}$ were dropped.  While our
1-point nulling works impressively well, the cut in multipole space is as, and
perhaps even more, impressive.  For example, to achieve the same statistical 
error as in 1-point nulling with one combination of shears dropped (error $\gtrsim 100\%$
bigger than the fiducial one) we can alternatively cut all multipoles at
$\ell>1000$, but then the bias with cutting in $\ell$ is more than two times smaller
than the bias with nulling! We have checked that these results are
qualitatively unchanged with a different choice of both the actual 
and the guessed bias.

\section{2-point nulling}
\label{sec:2pt}

Inspection of Table \ref{tab:degrade_1pt} shows that one problem with 1-point
nulling is that it does not have enough resolution in choosing how much of the
small-scale structure to null. For example, dropping the single combination of
shears that is most sensitive to the parameter $a$ increases the error in $w$
by about 150\%, which is too much to tolerate regardless of the success in
nulling out the unwanted bias in $P(k, z)$. The lack of ability to perform
nulling more gradually is not too surprising, as there are only $N_s$ linear
combinations to choose from.

Ideally we would like to have a better resolution in the eigenvalues
$\lambda_i$ and thus obtain a better leverage in controlling the cosmological
parameter degradations as a function of the nulling efficiency.  One way to
remedy that is to form linear combinations of the convergence power
spectra $C_{ij}^{\kappa}$, as there are $N\equiv N_s(N_s+1)/2$ of them
\begin{equation}
v_i\equiv  \sum_{j=1}^{N} \tilde{Q}_{ij}C_j^{\kappa} (\ell)
\label{eq:Ckappa_comb}
\end{equation}
\noindent where $\tilde{Q}_{ij}$ are coefficients (different from $Q_{ij}$ in
the previous section) and we denote a pair of redshift bins by a single
subscript; here $i$ and $j$ take values from $1$ to $N$.  The goal is to find
the linear combination(s) $v_i$ that produce the largest Fisher matrix element
$F_{aa}$.  While the 1-point procedure can be repeated, the optimization
problem, sadly, cannot be solved analytically as before because $\langle
C_j^{\kappa}\rangle\neq 0 $, and we needed that assumption for the expression
for $F_{aa}$, Eq.~(\ref{eq:Faa}), to take its simple form.  Therefore, we
choose to find the optimal combinations of the convergence power spectra by
brute force.  Using Powell's minimization method \cite{NR}, we find
the linear combination $v_1$ that has maximal $F_{aa}$.  Using the Gram-Schmidt
algorithm, we then project to the $(N-1)$ dimensional subspace, perform the
maximization again, and find the second most sensitive combination, $v_2$. And
so on, until we find all combinations of the power spectra ordered by their
sensitivity to the small-scale effect.  As with the 1-point function, we apply
this algorithm {\it at every $\ell$ separately}, where $\ell$ is the center of
the corresponding band power window in multipole space.

\begin{table}[!t]
\begin{tabular}{||c|c|c||c|c|c||}
\hline
\multicolumn{3}{||c||}{\rule[-2mm]{0mm}{5mm} 1-point Nulling} &
\multicolumn{3}{c||}{\rule[-2mm]{0mm}{5mm} Cutting above $\ell_{\rm max}$ }\\
\rule[-2mm]{0mm}{5mm}  
Skipped   & $\sigma(w)$  & $\delta(w)$ &
$\ell_{\rm max}$ & $\sigma(w)$  & $\delta(w)$
\\\hline\hline
\rule[-2mm]{0mm}{6mm} 0 & 0.054 & 0.085 & 10000 & 0.054 & 2.630 \\\hline
\rule[-2mm]{0mm}{6mm} 1 & 0.130 & 0.023 & 3000  & 0.073 & 0.101 \\\hline
\rule[-2mm]{0mm}{6mm} 2 & 0.291 & 0.003 & 1000  & 0.117 & 0.010 \\\hline
\end{tabular}
\caption{Nulling out shear combinations that are most sensitive
to $a$ vs.\ simply cutting measurements at some $\ell_{\rm max}$
for a 10-bin tomography case.  The nulled $\delta P(k)$ was assumed to go as $k^3$ (with
$k_*=10$h Mpc$^{-1}$), the actual one as $k^{3.5}$.  While nulling in principle
works well, it leads to degradation in statistical errors that are about 150\%
or larger, and therefore too large to tolerate.  On the other hand, simply
throwing out power beyond some $\ell_{\rm max}$ is surprisingly efficient; for
example, restricting to $\ell_{\rm max}=1000$ leads also to a similar $\gtrsim
100\%$ increase in $\sigma(w)$, but with the bias in $w$ that is
smaller than with 1-point nulling.  More importantly, cutting in $\ell$ can be
done gradually, while the 1-point nulling necessarily leads to jumpy
changes in the statistical error and bias.  }
\label{tab:degrade_1pt}
\end{table}

We indeed find that the resolution in nulling is greater with 2-pt nulling than
with 1-point; for example, throwing out the most sensitive combination
increases $\sigma(w)$ by about 40\%, and not 150\%. However, we found two
problems with 2-point nulling.  First, as before we find that simple cutting
in $\ell$ works about as well or better. In addition, we find that the
biases in cosmological parameters are substantially suppressed only if the
form of the bias is nearly precisely known, and not in more realistic
situations when the bias is guessed incorrectly. These conclusions are
unchanged even after trying different forms of $f(k)$, looking at different
cosmological parameters, different values of $k_*$, and different numbers of
survey redshift bins. Therefore, unless our 2-point procedure is substantially
improved in some way, 1-point nulling is the more effective approach.

\section{Cutting in wavenumber $k$}

\begin{figure*}[!t]
\vspace{-1cm}
\epsfig{file=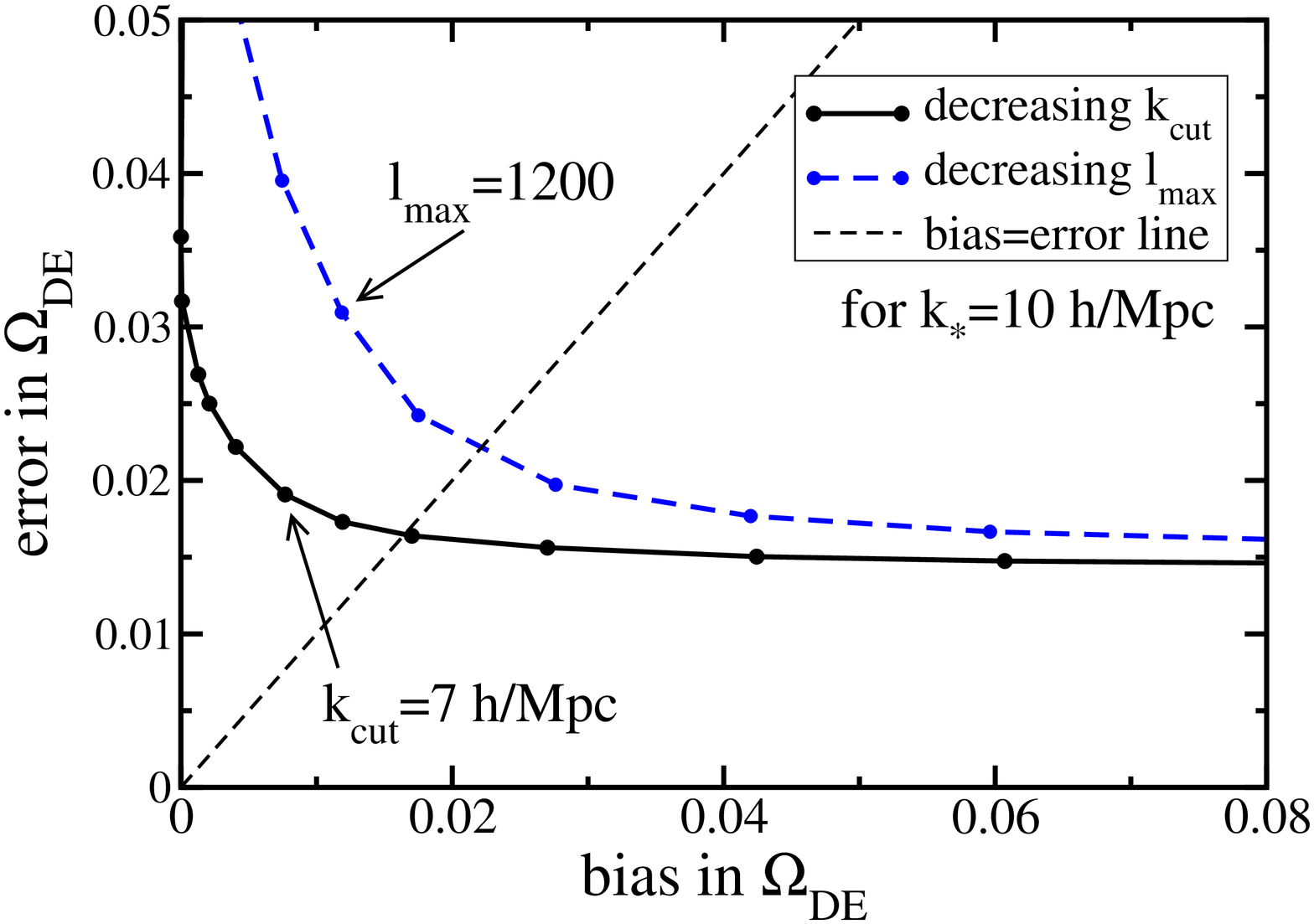,width=3.0in,height=3.4in,angle=-90}
\epsfig{file=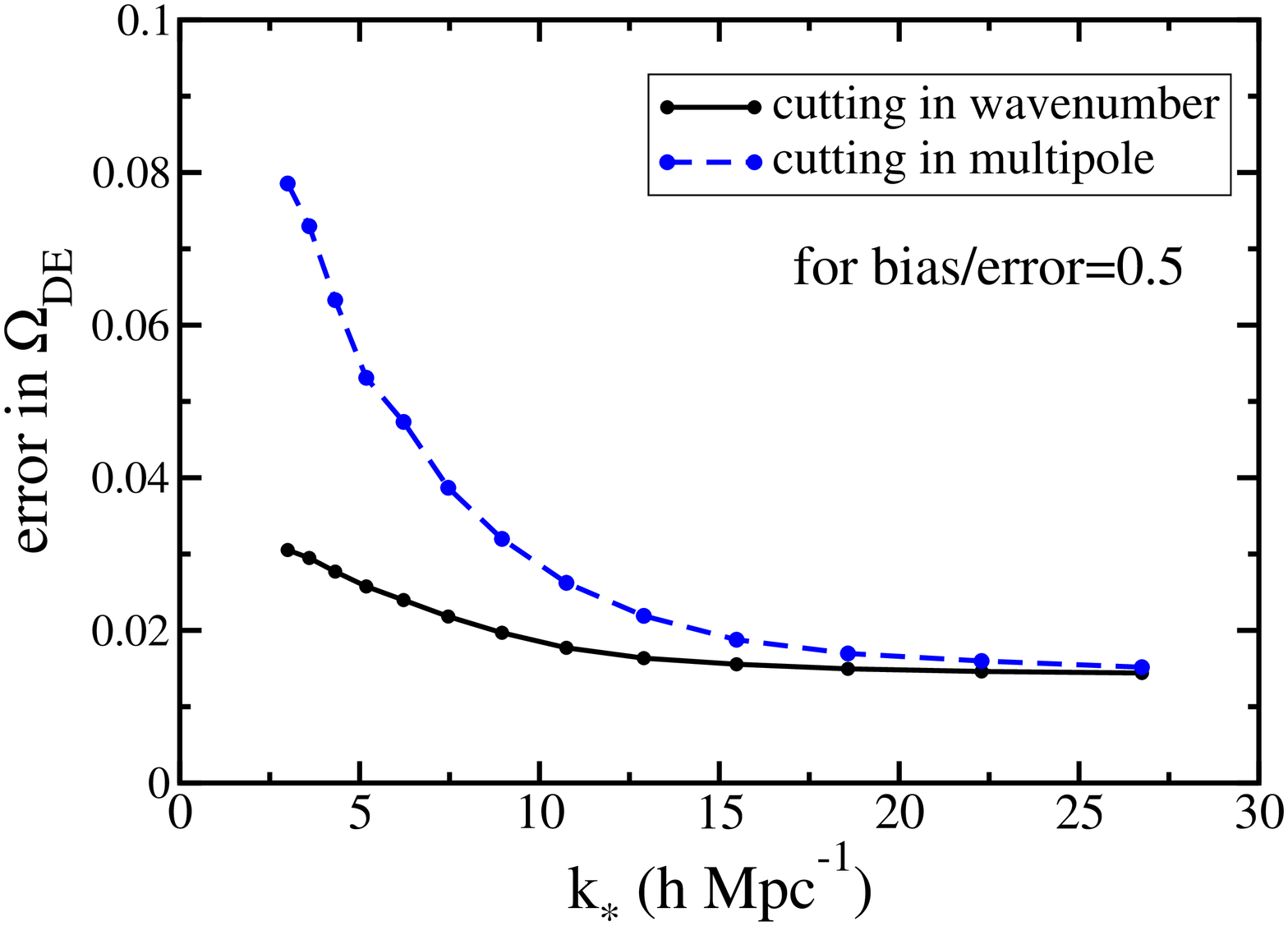,width=3.0in,height=3.4in,angle=-90}
\caption{Comparison of the cutting in wavenumber $k$ vs. the cutting in
multipole $\ell$.  Left panel: evolution of error vs.\ the (absolute) bias in
$\Omega_{\rm DE}$ with increased aggressiveness in cutting in $k$-space (solid
black curve) or $\ell$-space (dashed blue curve).  The cut is increasingly more
severe going from bottom right to top left. Here we use a 4-bin tomography case
with the small-scale effects going as $(k/k_*)^3$ with with $k_*=10\hmpcinv$.
For orientation, we show one value each of $\ell_{\rm max}$ and $k_{\rm cut}$
on the curves along which they vary; these are roughly the points where the
bias is reduced to a fraction ($\sim 0.5$) of the statistical error, and in this
case the total error with the $k$-cut is about 50\% smaller than that with the
$\ell$-cut. Right panel: statistical error in $\Omega_{\rm DE}$ after cutting
as a function of the scale at which the small-scale bias enters, $k_*$, for the
$k$ and $\ell$ cutting. With each method and for each value of $k_*$, just
enough cutting is performed so that the remaining bias in $\Omega_{\rm DE}$ is
equal to one half of its statistical error. Note that the two methods are
comparable if the bias enters at small scales (high $k_*$), but if the bias
affects much larger scales the $k$-cutting is clearly more effective as it leads to
a smaller final error in $\Omega_{\rm DE}$.  }
\label{fig:k_cut}
\end{figure*}

Another simple strategy is to simply throw out contributions to the band powers
coming from wavenumbers greater than some (comoving) cutoff $k_{\rm cut}$.  Assume that
the Limber's approximation integral over the large-scale structures is
discretized, with $N_p$ (rather than infinitely many) lens planes. Then,
assigning a single subscript $i$ for any of the $N$ cross-power
spectra, we have
\begin{equation}
C_i^{\kappa} (\ell)= \sum_{j=1}^{N_p} W_{ij} P(k_j, z_j)
\end{equation}
\noindent where $W$ are suitably defined weights that depend on the lensing
geometry and the distribution of galaxies and $P(k_j, z_j)$ is the matter power
spectrum evaluated at some wavenumber $k_j=\ell/r(z_j)$ and redshift $z_j$ of
the lens plane in question.  We now form a $N\times N_p$ matrix with the $(i,
j)$ entry equal to $W_{ij}P(k_j, z_j)$; we have $N=10$ (corresponding to $N_s=4$ redshift
bins) and $N_p=59$.  Then we simply transform this matrix to the lower triangular
form. The rows now represent linear combinations of the convergence power
spectra, and  new matrix entries are their weights. Of course, all $N$ rows
still contain the same information as before, as we simply performed a linear operation
on the $C^{\kappa}$s. However, by considering the first $M$ rows only
($M<N$), we can avoid using any information coming from wavenumbers greater than the
wavenumber on $(N_p-M)$th lens plane. As before, we repeat the procedure at
each multipole band centered at $\ell$. For simplicity we hold $k_{\rm cut}$
independent of $\ell$, and then higher $\ell$ implies higher $k_j$ for a fixed
lens plane --- therefore, more combinations will be cut at higher $\ell$, until
some multipole $\ell\simeq k_{\rm cut}r(z_{\rm max})$ where no information 
at $k>k_{\rm cut}$ will be left. Note too that, unlike the 1 and 2-point
nulling, cutting in $k$ does not require a guess of the offending
small-scale effect as all information beyond $k_{\rm cut}$, ``good'' and ``bad'', 
is thrown out.

To explore the efficacy of this type of nulling, we perform it for a variety of
values $k_{\rm cut}$, then compare to the ``gold standard'' of dropping
multipole power spectra above a fixed $\ell_{\rm max}$.  Overall, we find that
cutting in $k$ works better than cutting in $\ell$, as expected since our
small-scale contamination is defined in $k$. However, the two methods are
essentially equally powerful unless the small-scale effect enters at very large
scales (i.e.\ if $k_*$ is small), and in that case the $k$-cut can be
significantly more powerful; see Fig.~\ref{fig:k_cut}. This is easy to
understand as, from the relation $k\simeq \ell/r(z)$, we see that even if
stringent cutting in $\ell$ is performed, some amount of high $k$ modes, corresponding to
low redshift structures, will be left in the data --- and these high $k$ modes can be
extremely damaging if $k_*$ is small since the bias in the observables goes as
a power law in $k/k_*$. It is also interesting to note that cutting in $k$ is
equivalent to removing the low-redshift structures from the data --- in other
words, instead of cutting above some $k_{\rm cut}$, one could equivalently
choose not to use lens galaxies with (photometric) redshifts less than $z_{\rm
cut}$ where $r(z_{\rm cut})=\ell/k_{\rm cut}$.  Which of these two possibilities
is more feasible in practice is an interesting question, but outside the scope
of this work.

Current estimates indicate that the effective scale of various small-scale
effects is anywhere from a few to a few tens of $\hmpcinv$
(e.g. \cite{White_baryon, LosAlamos, Zhan_Knox}). If $k_*$ is closer to the
lower end of this range for any of these effects, cutting out the high
wavenumbers (or low-redshift structures) may prove extremely beneficial.

\section{Discussion and Conclusions}\label{sec:discussion} 

We have described and explored three different strategies for nulling out
small-scale information in weak lensing power spectrum measurements, with the
idea of protecting against uncertain, approximately known contamination while
preserving most of the sensitivity to cosmological parameters. 

The 1-point nulling tomography uses select linear combinations of shear in
different redshift bins so as to maximally reduce the survey's sensitivity to
small-scale bias.  The algorithm for achieving this is well known and
largely analytic, and it produces the desired linear combinations ordered by
how sensitive they are to the parameter $a$ describing the small-scale
bias. Figure \ref{fig:weights_1pt} shows that, as expected, the combination
that is most sensitive to the small-scale bias has weight at smaller scales
than the other combinations.  Since in a realistic situation we will only have
approximate understanding of the true bias, we test the performance of 1-point
nulling by assuming the fiducial bias is a power law in wavenumber, while the
fit bias, normalized by the parameter $a$, is a different power law.  We find
that even as just one linear combination of shear is dropped from the analysis
(the one most sensitive to the parameter $a$) the biases in cosmological
parameters are sharply reduced.  However, in that case the increase in
cosmological parameter errors is about 150\%.  Whether that is too large to
tolerate depends on the fiducial power of the survey and on the deleterious
effect of other, unrelated systematic errors.

The problem of large initial degradation can in principle be ameliorated by
2-point nulling, where we form linear combinations of the cross power spectra
(rather than shear). Because the number of cross power spectra is of order the
number of redshift bins {\it squared\/}, we have many more observables at our
disposal and can achieve a finer resolution in the amount of nulling performed,
and hence in cosmological parameter error degradation. Unfortunately 2-point
nulling does not have a nice analytic solution and linear combinations of
power spectra that are most sensitive to the small-scale bias need to be
found by brute force.  We find that the method works well in cases when the
functional form of the bias can be guessed accurately.
If the exact form for the bias is guessed incorrectly, however, 2-point
nulling produces biases in cosmological parameters that do not decrease
as the most sensitive combinations are dropped.
We conclude that 1-point nulling is more effective than the 2-point approach,
and this is unlikely to change unless a substantial improvement to our
2-point procedure is found in the future.

While both 1 and 2-point nulling are effective, we found that they are not
significantly better than the simplest strategy of dropping the highest
multipoles of the convergence power spectra.
[This may not be too surprising in retrospect: decreasing $\ell_{\rm max}$
{} from 10000 to 3000, for example, increases the cosmological parameter
error bars by only $\sim 30\%$ while having a tremendous impact in removing
a variety of small-scale contaminations --- in other words, the price to
pay with dropping the highest multipoles is relatively small, and this
procedure is already reasonably effective.]
This motivated us to explore yet another, but very different, method for
removing the small-scale biases.
We proposed a method of removing all information above some wavenumber
$k_{\rm cut}$ or, roughly equivalently, removing structures below some
redshift from the data.
This approach is different from the previous two in that the functional
form of the bias need not be known, and {\it all\/} information above
$k_{\rm cut}$, useful or not, is thrown out.
This works the best of all methods, leading to cosmological parameter errors
that are 10-50\% percent smaller than those with the $\ell$ space cut, and
by up to a factor of two if the small-scale effect enters at sufficiently
large scales. Given that weak lensing shear data is likely to be biased by
large nearby structures (e.g.~a large cluster of galaxies at low redshift),
some form of cutting in wavenumber will surely be beneficial to protect
against biases.

All of the aforementioned techniques require rough knowledge of the
cosmological model so that the fiducial convergence power spectra (i.e.\ those
not including the small-scale bias) are known to a good accuracy, and can be
manipulated to null out the unwanted effects. This should not be a problem in
the future, as the cosmological parameters are already reasonably well
determined with the combination of cosmic microwave background, type Ia
supernovae and large-scale structure constraints. However, systematic errors in
weak lensing measurements are a bigger concern. While the study in this paper
was precisely concerned with removing biases due to a class of systematics --
uncertain theoretical predictions on small scales -- there will be other
systematics in weak lensing measurements that can roughly be divided into
redshift errors and additive and multiplicative errors in measurements of shear
\cite{wlsys}. Those errors are likely to depend on numerous factors (e.g.\
observational season, galaxy morphology, etc.) and are uncertain at this time.
We do not believe that the systematics will significantly affect the nulling
techniques described in this paper since the systematics, provided they are
small, should be a lower-order contribution to our observables (which are
linear or quadratic in shear). Nevertheless, just as with all other
applications of weak lensing in the next generation experiments, incorporating
the systematics into the nulling tomography will be an important and
challenging task.

In conclusion, we strongly believe that the strategies that we described will
find application in weak lensing data mining. We only considered the
convergence power spectrum information here. Other aspects of weak lensing,
such as mapping of dark matter, depend more critically on the small-scale
information, and it might be that nulling-type techniques will reach their full
potential in precisely those applications.  It is likely that these interesting
possibilities will be considered in the future.

\vspace{0.5cm}
\section*{Acknowledgments}\vspace{-0.15cm}
We thank Wayne Hu for useful discussions and Eric Linder for comments on the manuscript. 
DH is supported by the NSF Astronomy and Astrophysics Postdoctoral Fellowship 
under Grant No.\ 0401066. MW is supported by NASA and the NSF.

\end{document}